\documentclass[conference]{IEEEtran}
\IEEEoverridecommandlockouts

\usepackage{cite}
\usepackage{amsmath,amssymb,amsfonts}
\usepackage{algorithmic}
\usepackage{graphicx}
\usepackage{textcomp}
\usepackage{xcolor}
\usepackage{placeins}
\usepackage{balance}

\newcommand{\addfig}[3]{
    \begin{figure}[hbt!]
        \centering
        \includegraphics[width=\linewidth,height=\textheight,keepaspectratio]{#3}
        \caption{\footnotesize{#2}}
        \label{#1}
    \end{figure}
}

\newcommand{\addfigcols}[3]{
    \begin{figure*}[hbt!]
        \centering
        \includegraphics[width=5in,height=\textheight,keepaspectratio]{#3}
        \caption{\footnotesize{#2}}
        \label{#1}
    \end{figure*}
}

\def\figref#1{Figure~\ref{#1}}
\def\tabref#1{Table~\ref{#1}}
\def\eqref#1{Equation~\ref{#1}}

\def\BibTeX{{\rm B\kern-.05em{\sc i\kern-.025em b}\kern-.08em
    T\kern-.1667em\lower.7ex\hbox{E}\kern-.125emX}}

\makeatletter
\newcommand{\linebreakand}{%
  \end{@IEEEauthorhalign}
  \hfill\mbox{}\par
  \mbox{}\hfill\begin{@IEEEauthorhalign}
}
\makeatother

\begin{document}

\title{Classic and Quantum Task-Based Intelligent Runtime for QIRs Running on Multiple QPUs}

\author{
\IEEEauthorblockN{Narasinga Rao Miniskar*}
\IEEEauthorblockA{
Oak Ridge National Laboratory\\
Oak Ridge, TN 37830, USA\\
miniskarnr@ornl.gov
}
\and
\IEEEauthorblockN{Elaine Wong*}
\IEEEauthorblockA{
Oak Ridge National Laboratory\\
Oak Ridge, TN 37830, USA\\
wongey@ornl.gov
}
\linebreakand
\IEEEauthorblockN{Vicente Leyton-Ortega}
\IEEEauthorblockA{
Oak Ridge National Laboratory\\
Oak Ridge, TN 37830, USA\\
leytonorteva@ornl.gov
} 
\and
\IEEEauthorblockN{Jeffrey S. Vetter}
\IEEEauthorblockA{
Oak Ridge National Laboratory\\
Oak Ridge, TN 37830, USA\\
vetter@ornl.gov
}
\and
\IEEEauthorblockN{Travis S. Humble}
\IEEEauthorblockA{
Oak Ridge National Laboratory\\
Oak Ridge, TN 37830, USA\\
humblets@ornl.gov
}
\thanks{*These authors contributed equally to this work and are designated as co-first authors.}
}

\maketitle

\begin{abstract}
High‑performance computing systems are rapidly evolving into heterogeneous platforms that fuse quantum accelerators with traditional classical processing units (CPUs) and graphical processing units (GPUs). This convergence calls for runtimes capable of managing both classical and quantum workloads in a unified manner. We introduce an intelligent, task‑based runtime that marries the Intelligent RuntIme System (IRIS) asynchronous scheduler with a quantum programming stack through the Quantum Intermediate Representation Execution Engine (QIR‑EE). Our design allows programs written in the quantum intermediate representation (QIR) to be dispatched concurrently to a variety of back‑ends, including multiple quantum simulators and nascent quantum processors, enabling genuine hybrid execution on a single node. To illustrate its practicality, we partition a 4‑qubit and 20-qubit circuit into three sub‑circuits using quantum circuit cutting via the QCut library. Each sub‑circuit is simulated independently by the QIR‑EE driver within IRIS, after which a classical post‑processing step merges the simulation results to recover the outcome of the original full‑circuit computation. This case study demonstrates how finer task granularity can enable the parallel execution and lower the simulation burden per quantum task while preserving overall accuracy, highlighting the feasibility of our hybrid approach.
\end{abstract}


\section{Introduction}
Hybrid quantum‑classical programming has emerged as a necessity for extracting speedups from nascent quantum processors, yet most quantum software stacks remain siloed, providing only a dedicated runtime for a single hardware platform.  While the quantum software community has been developing full‑stack solutions for dedicated quantum computing hardware, the essential concern of integrating such paradigms into existing software architectures, frameworks, and tools~\cite{pennylane2022,qiskit2024,pyquil2016} has largely remained unaddressed. The gap lies in the seamless orchestration of quantum and classical tasks within a single, scalable workflow.  To address this, we integrate the task‑based IRIS heterogeneous runtime~\cite{kim2024iris,miniskar2024iris} with the hardware‑agnostic QIR~\cite{qirspec} and its execution engine QIR‑EE~\cite{wong2025qiree}.  IRIS supplies a robust asynchronous scheduler \cite{johnston2024iris,miniskar2025iris,miniskar2023iris} that manages fine‑grained dependencies, dynamic device placement, and execution across CPUs, GPUs, and analogously, quantum processing units (QPUs) as either real quantum hardware or quantum simulators, while QIR‑EE offers a portable bridge between QIR code and diverse quantum back‑ends, including emerging QPUs.  By exposing QIR kernels as IRIS tasks, we enable concurrent scheduling of classical computations and quantum sub‑programs on the same node, thereby realizing a truly hybrid runtime that respects the data‑flow and synchronization requirements of both paradigms. 

The early efforts of Q‑IRIS (IRIS and QIR‑EE)~\cite{miniskar2026q} were limited to calling Python‑based Qiskit~\cite{qiskit2024} kernels for the given QIR programs and running them only on Qiskit's AER simulator, this paper presents a true classical‑quantum hybrid workflow. With recent extensions to QIR-EE integration in IRIS, the user can run the QIR programs on any QPU (real or simulator) through the eXtreme-scale ACCelerator (more commonly known as XACC) programming framework~\cite{xaccrepo}, which is connected to real hardware via vendor application programming interfaces (APIs) or Google's state simulator backeend QSIM~\cite{qsim}, which accessible directly from QIR-EE. Furthermore, IRIS has been extended to support classical post‑processing implemented in pure Python.  The extended framework executes these tasks on CPU cores through the IRIS heterogeneous‑memory model, which directly maps the outputs of QPU operations to NumPy objects. The quantum portion is executed on a variety of back‑ends, including local simulators and real QPU hardware (e.g., IONQ ion‑trap devices~\cite{ionq}), thereby enabling end‑to‑end hybrid applications that seamlessly couple classical processing with diverse quantum simulators and real quantum processors.

\addfigcols{fig:iris-background}{IRIS: Intelligent RuntIme System Design with QPUs. Proposed tightly coupled integration of QIR-EE in IRIS runtime is shown in green boxes.}{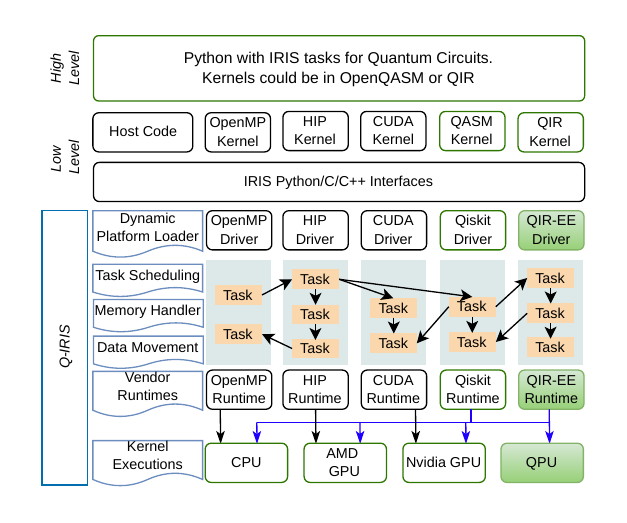}

\pagebreak
Our prototype demonstrates the practicality of this approach through a quantum circuit‑cutting experiment. A 4‑qubit and 20-qubit circuit are decomposed into three sub‑circuits, each simulated independently by QIR‑EE under IRIS’s orchestration.  A subsequent classical post‑processing task merges the simulation results to reconstruct the full circuit’s output, illustrating how task granularity reduces per‑quantum‑task load and improves throughput on early‑stage QPUs. This work not only showcases a portable hybrid programming model that scales across multiple QPUs but also highlights the potential for future research to extend the runtime’s capabilities—such as dynamic resource allocation and heterogeneous memory management—to fully exploit the evolving landscape of quantum‑classical high‑performance computing.

\section{Background}

\subsection{IRIS}
IRIS \cite{kim2024iris,miniskar2024iris} is a heterogeneous runtime system that brings together a broad spectrum of compute devices, including CPUs, Nvidia and AMD GPUs, Xilinx and Intel field programmable gate arrays (FPGAs), and Snapdragon Hexagon digital signal processors (DSPs), as shown in \figref{fig:iris-background}. Through a concise task‑based  API, developers can create and submit work items whose kernels are written in the appropriate programming model for each target: Compute Unified Device Architecture (CUDA) for Nvidia GPUs, Heterogeneous-computing Interface for Portability (HIP) for AMD GPUs, OpenCL for generic CPUs and GPUs, Xilinx high-level synthesis (HLS) C++ for FPGAs, and DSP‑specific C++ for Hexagon units. The runtime supplies multi-stream asynchronous capabilities~\cite{miniskar2025iris}, intelligent data orchestration~\cite{miniskar2023iris} automatically migrating data between host memory and the various accelerator memories to avoid programmer burden. Configurable scheduling policies allow the system to decide dynamically~\cite{johnston2024iris}  where each task should run, maximizing resource utilization. In addition, IRIS performs dynamic compilation so that kernels are lazily built and optimized at execution time. Together, these features streamline heterogeneous programming while preserving performance portability across the supported hardware platforms. The architecture is deliberately modular, enabling straightforward extension to new device types without disrupting existing workflows. As a result, developers can focus on algorithmic development rather than the intricacies of device‐specific code.

IRIS already serves as the execution engine for a number of high‑level programming frameworks such as OpenACC, SYCL (pronounced ``sickle"), Python, and Julia, demonstrating its versatility in real‑world applications. Its plug‑in‑style design allows the integration of additional compute units, exemplified by recent support for quantum processors via the QIR domain‑specific language \cite{miniskar2026q}, but it is limited to use it for single QPU (QSIM) due to QIR-EE thread safety limitations. In this work, we augment IRIS with complete QIR-EE quantum computing capabilities, thereby offering a unified platform where the same quantum code can be executed on virtual QPUs that simulate/emulate conventional CPUs and GPUs, and real dedicated quantum processing units. This extension leverages IRIS’s core mechanisms—data orchestration, dynamic scheduling, and runtime compilation—to adapt the workload to the strengths of each underlying device. By providing a single programming model across these diverse substrates, developers are freed from dealing with device‑specific programming paradigms.

\addfigcols{fig:iris-qiree-new}{IRIS with QIR-EE device (Q-IRIS) driver to connect to multiple QPUs through XACC. The QSIM simulator can be accessed separate from XACC and directly through QIR-EE via QIR-QSIM. The top green box illustrates the structure of the task graph in the form of a directed acyclic graph.}{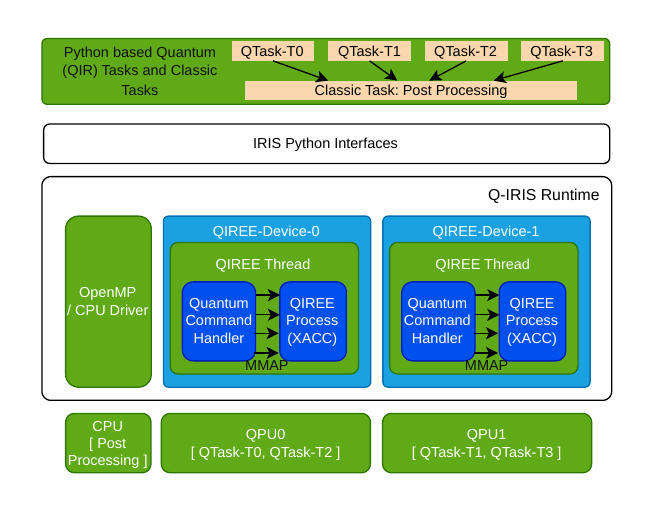}

\subsection{QIR-EE}
The Quantum Intermediate Representation Execution Engine (QIR‑EE) is a C++ library created by ORNL~\cite{qireerepo,wong2025qiree} that translates quantum programs written in the Quantum Intermediate Representation
(QIR)~\cite{qirspec} into executable code for both simulators and real hardware.  QIR itself was introduced to enable cross‑framework compatibility, allowing researchers to combine diverse programming languages with emerging
quantum devices.

Once compiled, the QIR‑EE binary accepts a QIR *.ll file, selects an accelerator target, and specifies the number of shots to run.  It then dispatches the program to the chosen backend, effectively serving as a
quantum runtime that can be plugged into the IRIS workflow as a QPU layer.

The engine relies on LLVM’s runtime infrastructure to process classical control flow while exposing hooks for binding backend‑specific operations.  Current support for simulators and hardware access is routed through the XACC
framework~\cite{xaccrepo}.  However, QIR‑EE presently lacks the ability to launch multiple QIR programs concurrently across several accelerators, a limitation that the IRIS integration seeks to address as a proof‑of‑concept for
asynchronous, heterogeneous execution.

\section{IRIS with QIR-EE for classic and quantum}

The core of our hybrid runtime is IRIS, which is extended to recognize QIR kernels as first‑class tasks.  In this model, a user submits a QIR program together with any classical compute kernels (for example, post‑processing) to the IRIS task graph.  The IRIS runtime resolves fine‑grained data dependencies, assigns each task to an appropriate backend—be it a CPU core, GPU, quantum simulator, or an emerging QPU—and orchestrates concurrent execution across all available devices.  By decoupling the task definition from the underlying hardware, the same application code can run on a purely classical cluster or a mixed classical/quantum node without modification.

We have extended IRIS versatile heterogeneous runtime to encompass quantum processors by employing a dedicated QIR-EE device driver, which is shown in \figref{fig:iris-qiree-new}. IRIS scheduler orchestrates quantum tasks such as QTask‑T0, QTask-T1, QTask-T2 and QTask‑T3 to run it parallel on multiple QPUs, alongside a classical post‑processing stage task to run it on the CPU, all within the same heterogeneous schedule. Because QIR-EE’s runtime is not thread‑safe, the driver spawns an isolated process that runs the QIR-EE engine together with the XACC back‑end; communication is carried out via a memory‑mapped (MMAP) region that shares the QIR LLVM module, control commands, and final expectation values. 

To bridge QIR code to the heterogeneous back‑ends, we have exposed C-APIs in QIR-EE and have written a QIR-EE driver in IRIS which uses these APIs to 1) load the QIR module from memory,
2) compile them on-the fly, 3) estimate the amount of memory size required for output expectation values of given QIR quantum circuit, 4) run the circuit either using QSIM or the XACC backend based on the task kernel configuration, and 5) extract the expectation value results into IRIS heterogeneous memory model. 

This tight integration removes the serial bottleneck that previously plagued QIR runtimes.  Because the driver can concurrently dispatch multiple QIR programs—each possibly accompanied by auxiliary classical tasks—to different
accelerators, the IRIS scheduler can now exploit the full spectrum of hardware resources in parallel.  The result is a seamless, high‑throughput workflow that blends quantum and classical computation while preserving the simplicity of the IRIS programming model.

\section{Results}

\begin{table*}[!h]
\centering
\parbox{\textwidth}{\centering
\caption{Circuit cutting with finite shot uncertainty propagation ($\sigma$) using Q-IRIS (1000 shots, 4 qubits in GHZ)}
\label{tab:simulator-experiments-sigma}
\footnotesize
\setlength{\tabcolsep}{6pt}
\renewcommand{\arraystretch}{1.1}

\begin{tabular}{lcccccccc}
\hline
\textbf{Backend Used}
& \boldmath$\langle ZZZZ \rangle$
& \textbf{$\sigma$}
& \textbf{Full (s)}
& \textbf{Create (s)}
& \textbf{Exec+Post (s)}
& \textbf{Retrieve (s)}\\
\hline \hline

XACC/AER              & 0.971745 & 0.038755 & 4.2048 & 0.1683 & 3.9710 & 0.0003\\
XACC/QPP              & 0.989037 & 0.038768 & 4.2044 & 0.1766 & 3.9595 & 0.0002\\
XACC/QSIM             & 0.983038 & 0.038759 & 4.1511 & 0.1784 & 3.9068 & 0.0002\\
XACC/IONQ:SIM.IDEAL   & 1.000000 & 0.044721 & 72.9430 & 0.1727 & 72.7037 & 0.0003\\
XACC/IONQ:SIM.ARIA-1  & 1.059612 & 0.044092 & 81.2316 & 0.1761 & 80.9893 & 0.0003\\
QIR-QSIM              & 0.989994 & 0.038759 & 3.6968  & 0.1936 & 3.4341 & 0.0002\\

\hline
\end{tabular}
}
\end{table*}

In our prototype, we demonstrate the practicality of this approach with a quantum circuit‑cutting experiment~\cite{miniskar2026q}. A 4‑qubit and 20-qubit Greenberger–Horne–Zeilinger (GHZ) circuit is partitioned into three sub‑circuits via a wire-cutting procedure described in~\cite{harada2024doubly} and briefly illustrated in Fig.~\ref{fig:qcut-fig-simple} (for the 4-qubit case).

\addfig{fig:qcut-fig-simple}{Illustration of how to partition a 4-qubit GHZ circuit using two wire cuts, creating three 2-qubit sub-circuits. The $|k\rangle, |s\rangle$ are initialized based on the measurement outcome (orange) in the previous circuit (depending on the measurement basis), and can be pre-computed. Their corresponding outputs $o_k,o_s$, together with outputs $y_1,y_2,y_3,y_4$ is used to construct the estimator for $\langle ZZZZ \rangle$ via the quasi-probability decomposition formula~\cite{harada2024doubly,miniskar2026q}. The 20-qubit circuit is partitioned similarly for $\langle Z^{\otimes 20}\rangle$ estimation.}{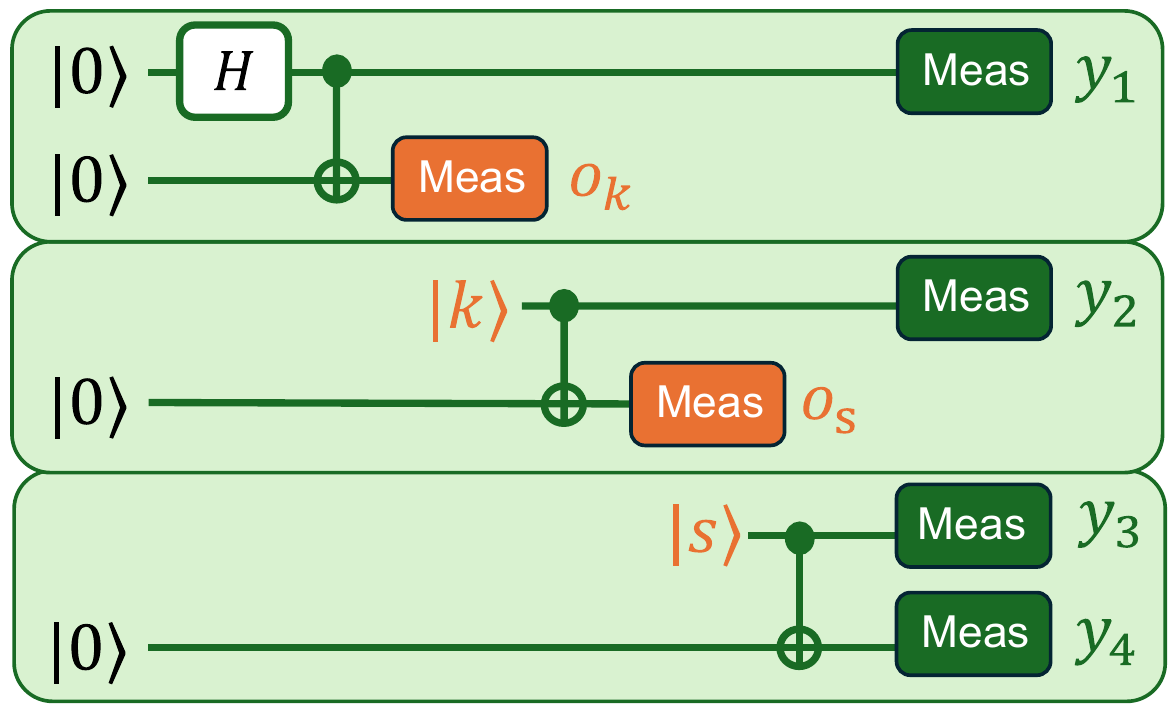}

Each sub‑circuit is dispatched as an independent IRIS task to a quantum simulator through the QIR-EE driver, which gives access directly to the Google's simulator QSIM, or to XACC, which provides hardware backend access. The overall workload is thereby reduced by a factor of more than three but increases the number of quantum tasks which can be run in parallel. Once all sub‑circuits finish, a classical post‑processing task aggregates the measurement outcomes to estimate the full circuit result. Per the algorithm consisting of two wire cuts, the resulting three sub-circuits produces 192 task kernels to be dispatched to a QPU backend~\cite{miniskar2026q, qcutrepo}. A post-processing task works on the measurement outcomes, represented as probability distributions of bitstring counts, which are then used to estimate the expectation value of an observable (whose theoretically known value can be used as ground-truth). This reconstruction is compiled into an estimation formula for $\langle ZZZZ\rangle$ via the quasi-probability decomposition:
\begin{eqnarray*}
\gamma^2 \sum_{k,s} \frac{|c_k|}{\gamma}\frac{|c_s|}{\gamma} \sum_{\substack{(y_1, ..., y_4),\\(o_s, o_k)}}\text{sgn}(c_k)\text{sgn}(c_s) o_k o_s f(y_1, y_2, y_3, y_4) \\ 
\times P[y_1, o_k | k] \cdot P[y_2, o_s | k] \cdot P[y_3, y_4 | s], 
\end{eqnarray*}
where $P[y_1, o_k | k]$, $P[y_2, o_s | k]$, and $P[y_3, y_4 | s]$ represent the probability distributions conditioned on indices $k,s$, with $o_{k}$ and $o_{s}$ being the measurements of ancillary qubits (see Fig.~\ref{fig:qcut-fig-simple}) and $\gamma$, $|c|$'s, $f$ representing overhead factors, weights, and a function that emulates the behavior of the coefficients resulting from the wire cuts, respectively.
This case study shows that finer task granularity lowers the per‑quantum‑task simulation burden while maintaining reasonable accuracy, validating the feasibility of our hybrid runtime and laying the groundwork for future extensions such as dynamic resource allocation and heterogeneous memory management. 

\tabref{tab:simulator-experiments-sigma} summarizes the performance of our Q‑IRIS framework when a 4‑qubit GHZ circuit is partitioned into three sub‑circuits (two cuts) which results in 192 2-qubit circuits (quantum tasks), later to be reassembled according to the formula for estimating the expectation value of the desired observable. All simulator backends produce expectation values for $\langle ZZZZ\rangle$ that are within $<$5\% of the ideal value~1, confirming the effectiveness of the cut‑and‑re‑assemble strategy in preserving circuit fidelity. Furthermore, we report $\sigma$'s as propagated shot noise from the three sub-circuits in our simulator experiments for the 4-qubit GHZ. These numbers would improve as we increase the number of shot counts to our desired accuracy, e.g., increasing number of shots to 4000 halves $\sigma$. We note that the propagated uncertainty is present because circuit cutting reconstructs the observable from many finite-shot fragment estimators, whose variances accumulate and are amplified by the reconstruction weights. In summary, the observed deviation in $\langle ZZZZ\rangle$ may reflect both statistical uncertainty from finite-shot reconstructed estimators and systematic effects (backend/noise/reconstruction bias), but our propagated error ($\sigma$) quantifies the statistical part only.

\tabref{tab:simulator-experiments-nosigma} shows some results from a vendor backend, provided by the Oak Ridge Leadership Computing Facility's (OLCF) Quantum Computing User Program (QCUP)~\cite{QCUP} via vendor APIs (currently integrated in XACC) for executing our quantum circuits. For testing our framework, we leveraged IONQ's trapped-ion  technology~\cite{ionq} and simulators. The IONQ's ion‑trap simulators (IDEAL and ARIA-1) incurs a slowdown ($\sim$72-81s) due to remote communications, while the real QPU QPU.FORTE-1 is orders of magnitude slower ($\sim$22,000s), which is due to significant wait time for physical hardware to be available for each batch that is sent remotely.

\begin{table*}[!ht]
\centering
\parbox{\textwidth}{\centering
\caption{Backend experiments on hardware and a state simulator (100 shots)}
\label{tab:simulator-experiments-nosigma}
\footnotesize
\setlength{\tabcolsep}{6pt}
\renewcommand{\arraystretch}{1.1}

\begin{tabular}{lccccccc}
\hline
\textbf{Backend Used}
& \boldmath$\langle Z^{\otimes n} \rangle$
& \textbf{Full (s)}
& \textbf{Create (s)}
& \textbf{Exec+Post (s)}
& \textbf{Retrieve (s)}
& \textbf{$n$}
& \textbf{\# Cut Circuits}\\
\hline\hline

XACC/IONQ:QPU.FORTE-1 & 0.769232 & 22061.0500 & 0.1400 & 22060.8600 & 0.0003 & 4 & 192\\
XACC/IONQ:QPU.FORTE-1 & 0.980198 & 407.0000   & 0.1400 & 406.8597   & 0.0003 & 4 & no cut\\
\hline\hline
QIR-QSIM              & 0.973800 & 1.0605     & 0.1757 & 0.8450     & 0.0002 & 20 & 192\\
QIR-QSIM              & 1.000000 & 82.74413   & 0.0014 & 82.6956    & 0.0026 & 20 & no cut\\

\hline
\end{tabular}
}
\end{table*}

In the real QPU case, we have observed a lower expectation value estimation as we didn't choose any error mitigation approach during the run. We also observe the effect of hardware and shot noise on the final outcome by running a no-cut version of the experiment. In contrast, the QIR-EE direct implementation of QSIM achieves the fastest run ($\sim$1s), demonstrating the advantage of lightweight simulation backends for small sub‑circuits without the need to rely on XACC.  When we do not apply cutting and run a full 20‑qubit GHZ on QIR-EE's QSIM, the wall‑time rises to $\sim$83s, illustrating that circuit cutting can drastically reduce the per‑task simulation load and improve overall throughput on early‑stage QPUs while still maintaining accuracy.

\section{Conclusion}
We presented a classic‑and‑quantum task‑based intelligent runtime that couples the IRIS runtime framework with QIR-EE, enabling QIR programs to be executed concurrently on multiple QPUs and classical accelerators. By modeling QIR kernels as IRIS tasks, the system achieves fine‑grained scheduling, dynamic device placement, and true hybrid execution on a single node. A 4‑qubit and 20-qubit circuit cut into three sub‑circuits, each running on both simulator QPUs and real QPUs (IONQ) independently, demonstrates that task granularity preserves accuracy while reducing per‑quantum load and improving throughput on early‑stage QPUs. These results confirm the practicality of a portable hybrid model that scales across diverse simulators and real quantum hardware. Future work will extend backend capability for Q-IRIS independent of XACC (in progress are Xanadu's suite of Lightning simulators~\cite{Lightning}), dynamic resource allocation, heterogeneous memory management, and multi‑node support to fully exploit the evolving quantum‑classical HPC ecosystem. 

\section*{Acknowledgment}
    This manuscript has been authored by UT-Battelle LLC under contract DE-AC05-00OR22725 with the US  Department of Energy (DOE). The US government retains and the publisher, by accepting the article for publication, acknowledges that the US government retains a nonexclusive, paid-up, irrevocable, worldwide license to publish or reproduce the published form of this manuscript, or allow others to do so, for US government purposes. DOE will provide public access to these results of federally sponsored research in accordance with the DOE Public Access Plan (https://www.energy.gov/doe-public-access-plan).
    
    This work was supported by the U.S. Department of Energy, Office of Science under Contract No. DE-AC05-00OR22725, with funding from the Office of Advanced Scientific Computing Research's Accelerated Research in Quantum Computing Program's Modular and Error-Aware Software Stack for Heterogeneous Quantum Computing Ecosystems (MACH-Q) project and resources of Oak Ridge National Laboratory's Experimental Computing Laboratory (ExCL).

    This research also used resources of the Oak Ridge Leadership Computing Facility, which is a DOE Office of Science User Facility supported under Contract DE-AC05-00OR22725. In particular, the Quantum Computing User Program (QCUP) run by OLCF enabled the use of IONQ's backends. The authors thank Michael Sandoval from ORNL and Heidi Nelson-Quillin from IONQ for their technical support. 

\balance    
\bibliographystyle{IEEEtran}
\bibliography{bibliography}

\begin{thebibliography}{10}
\providecommand{\url}[1]{#1}
\csname url@samestyle\endcsname
\providecommand{\newblock}{\relax}
\providecommand{\bibinfo}[2]{#2}
\providecommand{\BIBentrySTDinterwordspacing}{\spaceskip=0pt\relax}
\providecommand{\BIBentryALTinterwordstretchfactor}{4}
\providecommand{\BIBentryALTinterwordspacing}{\spaceskip=\fontdimen2\font plus
\BIBentryALTinterwordstretchfactor\fontdimen3\font minus \fontdimen4\font\relax}
\providecommand{\BIBforeignlanguage}[2]{{%
\expandafter\ifx\csname l@#1\endcsname\relax
\typeout{** WARNING: IEEEtran.bst: No hyphenation pattern has been}%
\typeout{** loaded for the language `#1'. Using the pattern for}%
\typeout{** the default language instead.}%
\else
\language=\csname l@#1\endcsname
\fi
#2}}
\providecommand{\BIBdecl}{\relax}
\BIBdecl

\bibitem{pennylane2022}
\BIBentryALTinterwordspacing
V.~Bergholm, J.~Izaac, M.~Schuld, and et~al., ``Pennylane: Automatic differentiation of hybrid quantum-classical computations,'' 2022. [Online]. Available: \url{https://arxiv.org/abs/1811.04968}
\BIBentrySTDinterwordspacing

\bibitem{qiskit2024}
A.~Javadi-Abhari, M.~Treinish, K.~Krsulich, C.~J. Wood, J.~Lishman, J.~Gacon, S.~Martiel, P.~D. Nation, L.~S. Bishop, A.~W. Cross, B.~R. Johnson, and J.~M. Gambetta, ``Quantum computing with {Q}iskit,'' 2024.

\bibitem{pyquil2016}
R.~S. Smith, M.~J. Curtis, and W.~J. Zeng, ``A practical quantum instruction set architecture,'' 2016.

\bibitem{kim2024iris}
J.~Kim, S.~Lee, B.~Johnston, and J.~S. Vetter, ``{IRIS}: {A} performance-portable framework for cross-platform heterogeneous computing,'' \emph{IEEE Transactions on Parallel and Distributed Systems}, 2024.

\bibitem{miniskar2024iris}
N.~R. Miniskar, S.~Lee, J.~Beau, A.~Young, M.~A.~H. Monil, P.~Valero-Lara, and J.~S. Vetter, ``{IRIS} reimagined: {A}dvancements in intelligent runtime system for task-based programming,'' in \emph{Workshop on Asynchronous Many-Task Systems and Applications}.\hskip 1em plus 0.5em minus 0.4em\relax Springer, 2024, pp. 46--58.

\bibitem{qirspec}
\BIBentryALTinterwordspacing
{QIR Alliance}, \emph{{QIR} Specification}, 2025, accessed: 2025-01-13. [Online]. Available: \url{https://github.com/qir-alliance/qir-spec}
\BIBentrySTDinterwordspacing

\bibitem{wong2025qiree}
E.~Wong, V.~Leyton-Ortega, D.~Claudino, S.~R. Johnson, A.~J. Adams, S.~Afrose, M.~Gowrishankar, A.~Cabrera, and T.~S. Humble, ``A cross-platform execution engine for the quantum intermediate representation,'' \emph{The Journal of Supercomputing}, vol.~81, no.~16, p. 1521, 2025.

\bibitem{johnston2024iris}
B.~Johnston, N.~R. Miniskar, A.~Young, M.~A.~H. Monil, S.~Lee, and J.~S. Vetter, ``{IRIS}: {E}xploring performance scaling of the intelligent runtime system and its dynamic scheduling policies,'' in \emph{2024 IEEE International Parallel and Distributed Processing Symposium Workshops (IPDPSW)}.\hskip 1em plus 0.5em minus 0.4em\relax IEEE, 2024, pp. 58--67.

\bibitem{miniskar2025iris}
N.~R. Miniskar, A.~R. Young, M.~A.~H. Monil, K.~Asifuzzaman, B.~Johnston, K.~Teranishi, and J.~S. Vetter, ``Iris-mash: Efficient multi-device asynchronous multi-stream heterogeneous computing,'' in \emph{Proceedings of the 54th International Conference on Parallel Processing}, 2025, pp. 764--773.

\bibitem{miniskar2023iris}
N.~R. Miniskar, M.~A.~H. Monil, P.~Valero-Lara, F.~Y. Liu, and J.~S. Vetter, ``{IRIS-DMEM}: {E}fficient memory management for heterogeneous computing,'' in \emph{2023 IEEE High Performance Extreme Computing Conference (HPEC)}.\hskip 1em plus 0.5em minus 0.4em\relax IEEE, 2023, pp. 1--7.

\bibitem{miniskar2026q}
N.~R. Miniskar, M.~A.~H. Monil, E.~Wong, V.~L. Ortega, J.~S. Vetter, S.~R. Johnson, and T.~Humble, ``Q-iris: The evolution of the iris task-based runtime to enable classical-quantum workflows,'' in \emph{Proceedings of the Supercomputing Asia and International Conference on High Performance Computing in Asia Pacific Region Workshops}, 2026, pp. 323--329.

\bibitem{xaccrepo}
{XACC Developers}, ``{XACC},'' \url{https://github.com/ORNL-QCI/xacc}, 2025, accessed: 2025-01-13.

\bibitem{qsim}
\BIBentryALTinterwordspacing
{Quantum AI Team and Collaborators}, ``{QSIM},'' Jun. 2025. [Online]. Available: \url{https://doi.org/10.5281/zenodo.4067237}
\BIBentrySTDinterwordspacing

\bibitem{ionq}
\BIBentryALTinterwordspacing
{IonQ Developers}, ``{IonQ},'' 2026, trapped-Ion Quantum Computing Company. Last accessed: March 11, 2026. [Online]. Available: \url{https://ionq.com}
\BIBentrySTDinterwordspacing

\bibitem{qireerepo}
\BIBentryALTinterwordspacing
{QIR-EE Developers}, ``{QIR-EE},'' [Computer Software] \url{https://doi.org/10.11578/qiree/dc.20250114.1}, 2025. [Online]. Available: \url{https://github.com/ORNL-QCI/qiree}
\BIBentrySTDinterwordspacing

\bibitem{harada2024doubly}
H.~Harada, K.~Wada, and N.~Yamamoto, ``Doubly optimal parallel wire cutting without ancilla qubits,'' \emph{PRX Quantum}, vol.~5, no.~4, p. 040308, 2024.

\bibitem{qcutrepo}
\BIBentryALTinterwordspacing
{QCut Developers}, ``{QCut},'' 2026, published as computer software. Last accessed: February 14, 2026. [Online]. Available: \url{https://github.com/JooNiv/QCut}
\BIBentrySTDinterwordspacing

\bibitem{QCUP}
{Oak Ridge Leadership Computing Facility}, ``{Quantum Computing User Program},'' \url{https://docs.olcf.ornl.gov/quantum/quantum\_access.html}, 2026, accessed: 2026-03-11.

\bibitem{Lightning}
{X}anadu, ``{Pennylane Lightning Simulator},'' \url{https://docs.pennylane.ai/projects/lightning/en/stable/index.html}, 2025, accessed: 2026-03-11.

\end{thebibliography}

\end{document}